\documentclass[aps,floatfix,pra,twocolumn]{revtex4}

\usepackage{graphicx}
\usepackage{graphics}
\usepackage{amssymb}
\usepackage{amsmath}
\usepackage{epsfig}
\usepackage{latexsym}

\newcommand{\ket}[1]{|#1\rangle}
\newcommand{\bra}[1]{\langle #1|}
\newcommand{\braket}[2]{\langle #1|#2\rangle}
\newcommand{\op}[1]{\hat{#1}}
\newcommand{\sx}{\hat{\sigma}_{x}}
\newcommand{\sy}{\hat{\sigma}_{y}}
\newcommand{\sz}{\hat{\sigma}_{z}}
\newcommand{\se}{\hat{\sigma}_{\epsilon}}
\newcommand{\st}{\hat{\sigma}_{\theta}}

\newcommand{\lx}{L_{x}}
\newcommand{\ly}{L_{y}}
\newcommand{\lz}{L_{z}}

\newcommand{\lpi}{\hat{L}_{\varpi}}

\newcommand{\eqrf}[1]{(\ref{#1})}
\newcommand{\hf}{\frac{1}{2}}

\newcommand{\bea}{\begin{eqnarray}}
\newcommand{\eea}{\end{eqnarray}}
\newcommand{\ra}{\rightarrow}

\newcommand{\ua}{\uparrow}
\newcommand{\da}{\downarrow}

\newcommand{\ba}{\begin{array}}
\newcommand{\ea}{\end{array}}
\newcommand{\qt}{q_\theta}
\newcommand{\qe}{q_\epsilon}

\begin{document}

\title{Entanglement and bifurcations in Jahn-Teller models}
\author{Andrew P. Hines}
\email{hines@physics.uq.edu.au}

\author{Christopher M. Dawson}
\author{Ross H. McKenzie}
\author{G.J. Milburn}
\affiliation{Centre for Quantum Computer Technology, School of Physical Sciences, The University of Queensland, St Lucia, QLD 4072, Australia}

\date{\today}
\begin{abstract}
We compare and contrast the entanglement in the ground state of two Jahn-Teller models. The $E\otimes\beta$ system models the coupling of a two-level electronic system, or qubit, to a single oscillator mode, while the $E\otimes\varepsilon$ models the qubit coupled to two independent, degenerate oscillator modes. In the absence of a transverse magnetic field applied to the qubit, both systems exhibit a degenerate ground state. Whereas there always exists a completely separable ground state in the $E\otimes\beta$ system, the ground states of the $E\otimes\varepsilon$ model always exhibit entanglement. For the $E\otimes\beta$ case we aim to clarify results from previous work, alluding to a link between the ground state entanglement characteristics  and a bifurcation of a fixed point in the classical analogue. In the $E\otimes\varepsilon$ case we make use of an ansatz for the ground state. We compare this ansatz to exact numerical calculations and use it to investigate how the entanglement is shared between the three system degrees of freedom.

\end{abstract}

\maketitle

\section{Introduction}

The burgeoning field of quantum information science has provided new tools with which to probe the characteristics of complex quantum many-body systems. More specifically, the study of the entanglement properties of systems is an active area of research, aimed at shedding new light on previously studied fundamental systems.

As such there has been numerous studies of the entanglement in the ground states of various systems (see \cite{HNO03,LRV03,CM03,Hin03,OAF02,ON02,Gun01,LEB03,HNO03a,VMC03} and reference therein). Of particular interest has been those systems which exhibit a quantum phase transition, where it has been demonstrated that the entanglement properties are connected with this critical phenomena \cite{ON02,OAF02,LRV03,VLR02}.

Another problem where an understanding of the entanglement properties offers a new perspective is in the study of decoherence. Any real-life quantum system interacts and becomes entangled with its environment, causing quantum superposition states to decohere into classical statistical mixtures. One way of studying the process of decoherence in open quantum systems is by the quantum environment and studying the now closed system-environment setup.

Probably the most well-known system-environment model is the {\it spin-boson model} \cite{LCD+87,Wei93}, which describes the interaction between a qubit (any two-level system) and an infinite collection of harmonic oscillators, modeling the environment. The entanglement between the qubit and its `environment' (the oscillators) in the ground state of this model was recently studied by Costi and McKenzie \cite{CM03} where a further link between entanglement and QPT's was established.

 As a way of investigating the decoherence induced by certain measurements, Levine and Muthukumar \cite{LM03} consider a model describing a qubit coupled now to a single environmental mode. This system is also known as the $E\otimes\beta$ Jahn-Teller model \cite{Eng72}. Levine and Muthukumar \cite{LM03} study the variation in the ground state entanglement with respect to the strength of the coupling between the qubit and the oscillator. In the massive limit ($m\to\infty$) of the oscillator, two parameter regions are identified, where the ground state is completely separable and where the qubit and oscillator are entangled. In this article we aim to clarify this result, in the light of previous results from the authors \cite{HMM03a} and Lambert, Emary, and Brandes \cite{LEB03}, regarding ground state entanglement and corresponding fixed point bifurcations in the classical analogue.

Following the natural progression from the single oscillator case, we consider the $E\otimes\varepsilon$ Jahn-Teller system, which describes the coupling of a qubit to two identical (uncoupled) oscillators. Jahn-Teller models are of great importance in the study of the geometry of molecular structure, in cases where the coupling between electronic and nuclear states cannot be ignored \footnote{When there is no coupling of electronic and nuclear states, the problem of molecular structure is greatly simplified by the so-called Born-Oppenheimer approximation}. The $E\otimes\varepsilon$ Jahn-Teller system describes the coupling between a doubly degenerate electronic state ($E$) and a doubly degenerate normal mode ($\varepsilon$). Such a model has been used to study the degree of electron-nuclear entanglement in molecular states \cite{Sjo00}.

In the case of both Jahn-Teller models considered here, when there is no transverse magnetics field applied to the qubit, the ground state has a two-fold degeneracy. This means that there are an infinite number of ground states, consisting of all possible superpositions of any two orthogonal ground states. Not all such ground states will necessarily contain the same amount of entanglement. To obtain a complete picture of the ground state entanglement, one has to consider the entanglement in all possible ground states.

Often, it is the case that there simultaneously exists ground states with maximal entanglement, and completely separable ground states. Certainly, it can be shown that for a system of two qubits, if there are two orthogonal, maximally entangled ground states, then an equal superposition of the two is completely separable. This is the case for the $E\otimes\beta$ model, where, irrespective of the strength of the coupling, there is always a ground state which contains no entanglement. However, the $E\otimes\varepsilon$ model exhibits the intriguing property that for all ground states when the coupling is greater than zero, the qubit is entangled with the oscillators. While in the limit of large coupling there are ground states with maximal qubit-oscillators entanglement, we show that the entanglement in all ground states is always bounded below by some non-zero value.

We begin with the $E\otimes\beta$ model, by analyzing the corresponding classical model before considering the entanglement in the ground state. This is followed by the same analysis for the $E\otimes\varepsilon$ model and a comparison of the ground state entanglement characteristics of the two.

\section{$E\otimes\beta$ : a qubit coupled with a single oscillator mode}

The $E\otimes\beta$ is the mathematically simplest Jahn-Teller effect, and occurs where a doubly degenerate state (the qubit) becomes coupled by a single boson mode (the oscillator). The entanglement characteristics of such a model system have been recently studied by Levine and Muthukumar \cite{LM03}, where they considered a qubit coupled to a single harmonic oscillator described by the Hamiltonian

\begin{equation}
H = \Delta\sx + L\frac{1}{\sqrt{2m\omega}}\left(a+a^{\dagger}\right)\sz + \omega a^{\dagger}a,
\end{equation}

where $\omega$ is the natural frequency of the oscillator, $L$ is the coupling strength and $\Delta$ is the strength of the transverse magnetic field acting perpendicular to the coupling, all of which are in units such that $\hbar = 1$ (for the rest of the paper we assume unit mass, $m=1$). This Hamiltonian can also be written in terms of the position coordinate, $q$ of the oscillator, as
\begin{equation}\label{hamqbosc}
H = \Delta\sx + L\op{q}\sz -\frac{1}{2}\left(\frac{\partial^{2}}{\partial \op{q}^{2}} - \omega^{2}\op{q}^{2}\right).
\end{equation}

 This system is a simpler version of that studied by Emary and Brandes \cite{EB03}, who considered a collection of $N$ two-level atoms, modeled as a single collective spin, interacting with a single bosonic mode via a dipole interaction - the so-called Dicke Hamiltonian.
In their analysis based on functional integrals, Levine and Muthukumar \cite{LM03} identified a critical parameter value corresponding to a qualitative change in the ground state of the system. In the next section we consider the analogue classical system and derive this critical parameter via a simple analysis of the dynamical fixed points.

\subsection{Classical analogue and bifurcations}\label{classical}

Here we clarify that the critical parameter found by Levine and Muthukumar \cite{LM03} corresponds to a bifurcation of the fixed points \cite{Gle94} in the corresponding classical system.

Letting $q$ and $p$ be the classical position and momentum coordinates of the oscillator, and $\lx,\ly$ and $\lz$ the spin coordinates of the spinning top (the classical analogue of the qubit), the equations of motion are found to be
\begin{subequations}
\bea
\dot{q} &=& p,\label{ceq1}\\
\dot{p} &=& -L\lz - \omega^2q,\\
\dot{\lx} &=& -L q \ly,\\
\dot{\ly} &=& -\Delta\lz + L q\lx,\\
\dot{\lz} &=& \Delta\ly,\label{ceql}\
\eea
\end{subequations}
with the spherical constraint, $\lx^{2}+\ly^{2}+\lz^{2}=1$.

 Solving the above equations set to zero yields the fixed points of the system. It is simple to see that there exists two fixed points for all parameter values, at
\begin{equation}\label{ofp}
\lx=\pm 1,\lz=\ly=q=p=0,
\end{equation}
\noindent and for $L^{2} > \Delta\omega^{2}$ there exists a further four fixed points, located at
\begin{equation}\label{efp}
\lx = \pm\frac{\Delta\omega^{2}}{L^{2}},\hspace{.1cm} \lz=\pm\sqrt{1-\left(\frac{\Delta\omega^{2}}{L^{2}}\right)^{2}}, q = -\frac{L}{\omega^{2}}\lz,
\end{equation}
with $\ly=p=0$. Stability analysis of the fixed points shows that the original fixed points \eqrf{ofp} are stable for $L^2\leq\Delta\omega^{2}$, then lose their stability above this critical point, whilst the emergent fixed points are stable. The situation where a solitary fixed point becomes unstable and two new, stable fixed points emerge at some critical parameter value is called a {\it supercritical pitchfork} bifurcation. The bifurcation point, $L^2\leq\Delta\omega^{2}$, corresponds to the critical parameter values identified in Ref. \cite{LM03}.

 The bifurcation implies that above the critical point, the energy is minimized by assuming a non-zero value of the oscillator displacement, $x=\pm\frac{L}{\omega^{2}}\lz$, and the spin is now localized with a non-zero $\lz$.

  In a recent paper we studied this type of bifurcation and its relationship to entanglement \cite{HMM03a}. Lambert, Emary and Brandes \cite{LEB03} studied the entanglement in the more generalized system of a collection of $N$ qubits  coupled to a single oscillator. Since the qubits are all identically coupled to the oscillator mode, they can be modeled as a single qu{\it dit}, meaning this system has the same classical analogue as described in Sec.. \ref{classical}, exhibiting the same bifurcation. In the next section we study the qubit-oscillator entanglement in the ground state of Hamiltonian \eqrf{hamqbosc}, and finish by discussing the model of Lambert {\it et al} \cite{LEB03}.

\subsection{Entanglement in the ground state}

In their study of the characteristics of the ground state entanglement between the qubit and oscillator, Levine and Muthukumar \cite{LM03} focus on the determination of specific correlation functions via functional integrals with the characteristics of these functions being indicative of entanglement. We focus solely on a quantitative study of the entanglement, employing the canonical measure of bipartite entanglement, the entropy of entanglement, which is the von Neumann entropy of the reduced density operator, $\rho$ of the qubit i.e.,
\begin{equation}\label{VN}
S(\rho) = \rho \log_2 \rho.
\end{equation}
 To begin our study of the ground state entanglement, we consider the case where there is no transverse magnetic field applied to the qubit (i.e., $\Delta =0$).

\subsubsection{$\Delta=0$}
In the case of Hamiltonian \eqrf{hamqbosc} with $\Delta=0$ the eigenstate problem is exactly solvable \cite{Mah90}. Each energy eigenstate is two-fold degenerate, spanned by the (orthogonal) states
\bea
\braket{q}{\psi^{R}_{n}}&=\chi_n \left(q-\frac{L}{\omega^2}\right)\ket{\downarrow}&=\chi_n^R(q)\ket{\downarrow}\label{gsb1}\\
\braket{q}{\psi^{L}_{n}}&=\chi_n \left(q+\frac{L}{\omega^2}\right)\ket{\uparrow}&=\chi_n^L(q)\ket{\uparrow}\label{gsb2}
\eea
with energies,
\begin{equation}\label{energies}
E_n = \omega n -\frac{L^2}{2\omega^2},
\end{equation}
where $\chi_n (q)$ is the $n^{th}$ linear harmonic oscillator wavefunction. We see that $\ket{\psi^{L}_{n}}$ and $\ket{\psi^{R}_{n}}$ correspond to states localized in the left and right displaced harmonic wells, respectively. Note the correspondence with the fixed points derived earlier (\ref{ofp},\ref{efp}).
For the ground state, we have
\begin{equation}
\chi_0 (q) = \left(\frac{\omega}{\pi}\right)^{\frac{1}{4}}e^{-\frac{\omega}{2} q^2}.
\end{equation}
\noindent From the degeneracy, a general ground state can be written as any superposition of the states \eqrf{gsb1},\eqrf{gsb2}
\begin{equation}\label{supergs}
\ket{\psi_0} = c_1\ket{\psi^{L}_{0}}+ c_2 e^{i\gamma}\ket{\psi^{R}_{0}},
\end{equation}
with $c_1^2+c_2^2 = 1$. A general density operator describing the ground state is thus
\bea
\rho& = & c_1^2 \chi_0^L(q)^2\ket{\downarrow}\bra{\downarrow}+ c_1c_2\chi_0^L(q)\chi_0^R(q)\nonumber\\ & &\left(e^{-i\gamma}\ket{\downarrow}\bra{\uparrow}+ e^{i\gamma}\ket{\uparrow}\bra{\downarrow}\right)\nonumber\\ & &+c_2^2\chi_0^R(q)^2\ket{\uparrow}\bra{\uparrow}\nonumber.
\eea

\begin{figure}[th]
\begin{center}
\scalebox{.65}{\includegraphics{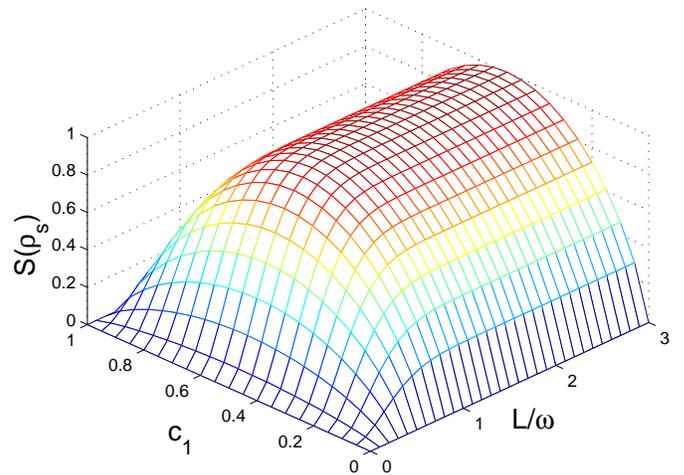}}
\end{center}
\caption{Entanglement in the ground state of the $E\otimes\beta$ system for different superpositions of the degenerate ground state (defined by Eq. \eqrf{supergs}) for increasing qubit-oscillator coupling, $L/\omega$.}
\label{qbosc}
\end{figure}

Tracing out the oscillator degree of freedom, the reduced density operator, $\rho_s$ is
\begin{equation}
\rho_s = \frac{1}{2} \left[ \begin{array}{cc}
    1 & c_1 c_2 e^{\alpha-i\gamma }   \\
    c_1 c_2e^{\alpha+i\gamma} & 1
\end{array} \right] \\
\end{equation}
where $\alpha = 2(L/\omega^2)^2$. This density operator allows the entropy of entanglement of the ground state to be determined as a function of $c_1$ and the coupling $L/\omega$ (it is independent of the phase $\gamma$) and is shown in figure \ref{qbosc}.

\begin{figure}[th]
\begin{center}
\scalebox{.45}{\includegraphics{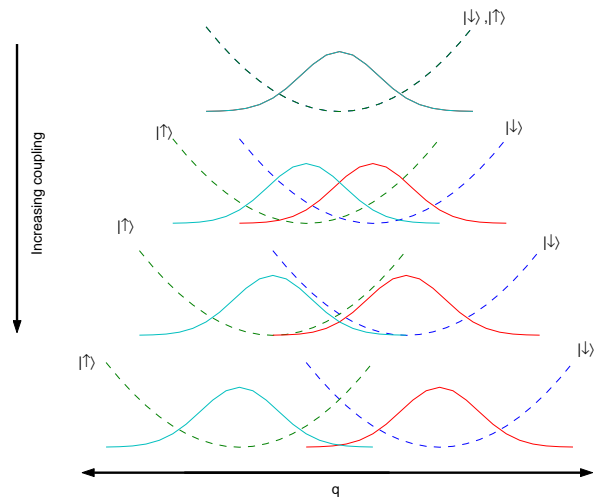}}
\end{center}
\caption{The coupling of the oscillator displacement to the spin acts to split the oscillator potential in two. The ground state is then either localized in one of the two potential wells - states $\ket{\psi^{R}_{0}},\ket{\psi^{L}_0}$ or a superposition of the two. As the coupling increases, the spatial separation of the two states increases. In turn, the overlap of the state decreases and the entanglement increases. The above corresponds to an equal superposition, which achieves the maximum entanglement.}
\label{diagram}
\end{figure}

Note there are two degenerate ground states ($\ket{\psi^{L}_{0}}$ and $\ket{\psi^{R}_{0}}$), where the qubit is never entangled with the oscillator, regardless of the coupling strength. For all superpositions of the two degenerate states, the entanglement increases, as the coupling, and hence the spatial separation of the two states increases (see figure \ref{diagram}). Maximum entanglement is achieved for an equal superposition.

\subsubsection{$\Delta \neq 0$}

The addition of the $\Delta \sx$ term to the Hamiltonian means that the eigenvalue problem is no longer exactly solvable so the ground state must be analyzed numerically (see appendix \ref{basis}).

\begin{figure}[t!]
\begin{center}
\scalebox{.45}{\includegraphics{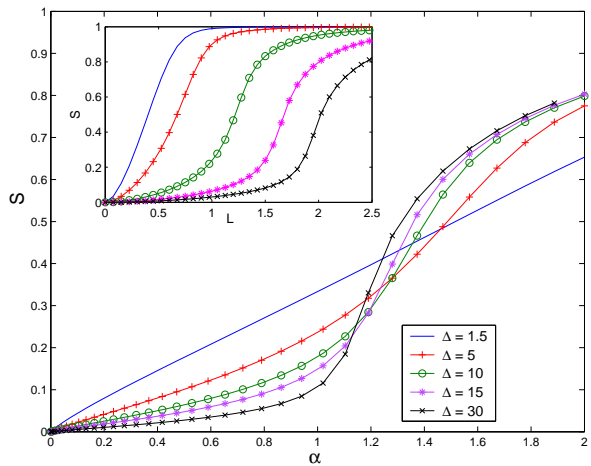}}
\end{center}
\caption{Entanglement in the ground state of the $E\otimes\beta$ system (with an applied transverse magnetic field) with respect to $\alpha = L^2 / \omega^2\Delta$ for various $\Delta$ (with $\omega = 1$). The inset is the same entanglement results, but with respect to solely the coupling strength, $L$. In all cases, the entanglement becomes maximal as $L \rightarrow \infty$. As the ratio of $\Delta/\omega$ increases, the distinction between the ``separable'' and ``entangled'' regions identified my Levine and Muthukumar \cite{LM03}, on either side of the classical bifurcation $\left(\alpha = 1\right)$, become apparent.}
\label{entsosc}
\end{figure}

 The $\Delta \sx$ term breaks the original degeneracy and forces the ground state to exhibit a superposition between the `up' and `down' spin states,  resulting in an entangled ground state for all $\Delta >0$ (see figure \ref{entsosc}). However, a non-zero $\Delta$ means the oscillator potential can no longer be viewed as two, spatially separate harmonic wells each corresponding to either of the two orthogonal states of the spin (`up' and `down') as shown in figure \ref{diagram}. Instead, the two separated wells now each correspond to some superposition of the spin states. This effect results in a decrease in the entanglement between the qubit and the oscillator (as evidenced in figure \ref{entsosc}).
In the limit of large $L$, the non-zero $\Delta$ ensures the ground state approaches a maximally entangled, equal superposition of the now far spatially separated states \eqrf{gsb1} and \eqrf{gsb2}.

Levine and Muthukumar focus on the entanglement in the ground state in the {\it massive} limit of the oscillator. More rigourously, this is defined as the limit of $m\rightarrow \infty$ while keeping $m\omega^2$ constant. Alternatively, this is equivalent to $\Delta /\omega \rightarrow \infty$. This is the limit of the quantized harmonic oscillator approaching its (continuous) classical counterpart. In Ref. \cite{LM03} it is argued that in this massive (classical) limit of the oscillator, the onset of entanglement in the ground state becomes discontinuous with respect to the parameter $\alpha = L^2 / \Delta m \omega^2$. This led Levine and Muthukumar to the identification of ``separable'' and``entangled'' parameter regions for the ground state - below the critical parameter, $\alpha_c = 1$, the ground state is separable and at $\alpha_c$ there is a discontinuous change in the ground state, whereby it becomes entangled.

In stating that the onset of entanglement becomes discontinuous, it is assumed that Levine and Muthukumar mean that the ground state entanglement with respect to the parameter $\alpha$ becomes non-analytic at the critical point. This is not surprising, since in the classical limit of the oscillator, the ground state does not change smoothly with respect to $\alpha$ at the critical $\alpha_c$.

In the classical limit, the ground state corresponds to the bifurcating fixed point, ($L_x = 1, L_z = q = 0$ for $\alpha < 1$) identified in Sec. \ref{classical}. As the oscillator behaves more classically, the change in the ground state with respect to $\alpha$ becomes non-analytic at $\alpha_c$. Due to the pitchfork nature of the bifurcation, the ground state transforms from the oscillator state localized around the single fixed point, to a superposition between the two emergent fixed points, as is passes through the bifurcation, i.e. $\langle q\rangle=0$ for $\alpha_c\leq 1$ while $\langle q\rangle=\frac{L}{\omega}\langle \sz\rangle = \pm q_0$ for $\alpha_c> 1$, where $\langle \sz\rangle\neq 0$. This is not the only model system where such a bifurcation can be used to infer an understanding of the entanglement properties of the ground state.

The system considered by Lambert, Emary and Brandes \cite{LEB03} describing the interaction of $N$ qubits with a single bosonic mode \cite{LEB03} (known as the Dicke model), undergoes a quantum phase transition in the $N\rightarrow\infty$ limit, at a critical value of the coupling, $L = L_c$.  Here the entanglement between the $N$-qubit ensemble and the field in the ground state, with respect to the coupling strength, $L$ was considered. It was demonstrated that the entanglement obtained its maximal value corresponding to the critical coupling. More interestingly, the entanglement goes to infinity and becomes discontinuous in the $N\rightarrow\infty$ limit.

The classical analogue of the Dicke model is identical to that defined in Sec. \ref{classical} with the critical coupling corresponding to the bifurcation in the classical analogue.

 In Ref. \cite{HMM03a}, we demonstrated that for a system of coupled giant spins whose classical analogue exhibits the same bifurcation, the entanglement between the spins with respect to the coupling strength is peaked at a coupling strength corresponding to the bifurcation. In the limit of infinite angular momentum, the maximum entanglement goes to infinity at this critical point.

In all three cases described above characteristics of the entanglement can be understood by considering the fixed-point bifurcation in the classical system.

We now take the next logical step and study the ground state entanglement in a system of a qubit coupled to {\it two} oscillators.

\section{$E\otimes\varepsilon$ : qubit with two degenerate oscillator modes}

The $E\otimes \varepsilon$ Jahn-Teller System models the interaction between a doubly degenerate electronic state ($E$) and a doubly degenerate normal mode ($\varepsilon$) \cite{Sjo00}. This is analogous to a qubit coupled to two harmonic oscillators. Following the notation of Englmann, the Hamiltonian modeling this system is defined as \cite{Eng72}

\begin{multline}
\label{englman::hamiltonian}
H = \hf\hbar\omega\left(\qe^{2}+\qt^2-\frac{\partial^2}{\partial \qt^{2}}-\frac{\partial\qe^2}{\partial \qe^{2}}\right)\\ + \hf L\left(\qt\sigma_\theta + \qe\sigma_\epsilon\right)
\end{multline}
\noindent where $\omega$ is the natural frequency of the identical oscillators and $L$ is the vibronic coupling strength (all in units of $\hbar$). In terms of the basis states of the qubit (or the electronic doublet), denoted
\bea
\ket{\da}= \left(\ba{c} 1 \\
            0 \ea\right)&,&\ket{\ua}=\left(\ba{c} 0 \\
            1 \ea\right),
\eea
the spin operators are defined as

\bea
\sigma_\theta= \left(\ba{cc} -1&0 \\
            0&1 \ea\right)&,&\sigma_\epsilon=\left(\ba{cc} 0&1 \\
            1&0 \ea\right).
\eea
Defining the usual oscillator mode creation and annihilation operators via
\bea
\qt &=& \frac{1}{\sqrt{2}}\left(a+a^\dagger\right)\\
p_\theta &=& \frac{i}{\hbar\sqrt{2}}\left(a^\dagger-a\right)\\
\qe &=& \frac{1}{\sqrt{2}}\left(b+b^\dagger\right)\\
p_\epsilon &=& \frac{i}{\hbar\sqrt{2}}\left(b^\dagger-b\right),
\eea
\noindent where $p_\theta = i\hbar\partial_\theta$ and $p_\epsilon = i\hbar\partial_\epsilon$ allows the Hamiltonian \eqrf{englman::hamiltonian} to be written as
\begin{multline}\label{var_englman}
H = \hbar\omega \left( a^\dagger a + b^\dagger b +1 \right)\\+\frac{L}{2\sqrt{2}}\left[\left(a+a^\dagger\right)\sigma_\theta + \left(b+b^\dagger\right)\sigma_\epsilon \right]
\end{multline}

\begin{figure}[ht]
\begin{center}
\scalebox{.7}{\includegraphics{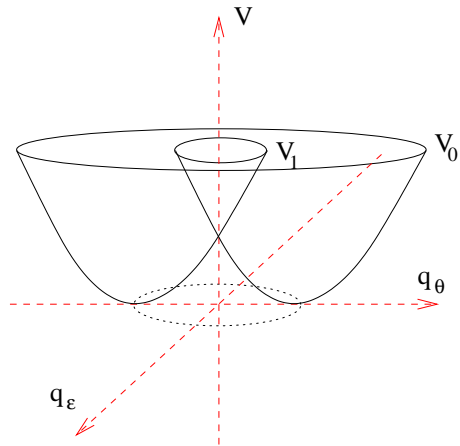}}
\end{center}
\caption{The Mexican-hat shaped potential. $V_{0}$ and $V_{1}$ correspond to the parabolic potentials of the individual harmonic oscillators, displaced from the origin by their coupling to the qubit. }
\label{mh}
\end{figure}

The adiabatic potential for this Hamiltonian has the `Mexican-hat' shape, as in figure \ref{mh}. Like the single oscillator case, the coupling of the qubit to the two, orthogonal oscillators, results in a splitting of the no parabolic potential in the two spatial oscillator dimensions.

\subsection{Conserved Quantity}

The total angular momentum of the system $\op{\mathbf{J}}$, is the sum of the orbital angular momentum $\op{\mathbf{L}}$ (contributed by the harmonic oscillators) and the spin angular momentum $\op{\boldsymbol{\sigma}}$ (contributed by the qubit) i.e. $\op{\mathbf{J}} = \op{\mathbf{L}}+\op{\boldsymbol{\sigma}}.$

Defining the direction $q_\varpi$, as that perpendicular to $\qt$ and $\qe$, it is possible to show that $$\op{J}_\varpi = \op{L}_\varpi+\sigma_\varpi $$ - the total angular momentum in the $\varpi$-direction - is a constant of the motion. Firstly, define
\begin{equation}
\op{\sigma}_\varpi = \left(\begin{array}{cc}
                0  & i\\
                -i & 0
              \end{array}\right)
\end{equation}
such that $\op{\sigma}_\epsilon,\op{\sigma}_\varpi,\st$ correspond to the Pauli's $\sx,\sy,\sz$ respectively and
\begin{equation}
\lpi = \op{q}_\theta\op{p}_\epsilon-\op{q}_\epsilon\op{p}_\theta=-i\hbar\frac{\partial}{\partial\phi}.
\end{equation}

Starting with the generic commutation relation relating position, $\op{q}$ and momentum $\op{p}$, $\left[\op{q},\op{p}\right]=i\hbar$, we now note the following commutation relations for $\lpi$ relating to the relevant terms in the Hamiltonian:
\bea
\left[\lpi,\op{q}_\theta\right] = i\op{q}_\epsilon&,&\left[\lpi,\op{q}_\epsilon\right]=-i\op{q}_\theta,\nonumber\\\left[\lpi,\op{q}_\theta^{2}\right] = 2i\op{q}_\epsilon\op{q}_\theta&,&
\left[\lpi,\op{p}_\theta^{2}\right] = 2i\op{p}_\epsilon\op{p}_\theta,\nonumber\\
\left[\lpi,\op{p}_\epsilon^{2}\right] = -2i\op{p}_\epsilon\op{p}_\theta&,&
\left[\lpi,\op{p}_\epsilon{2}\right] = -2i\op{p}_\epsilon\op{p}_\theta.\nonumber
\eea

Together with the Pauli spin operator commutation relations,
\begin{equation}
\left[\se,\hat{\sigma}_\varpi\right] = i\st \qquad \text{(and cyclic permutations)}
\end{equation}
 it is simple to see that $\left[ \op{J}_{\varpi},\op{H}\right] = 0$, so $\op{J}_{\varpi}$ is a constant of the motion. Note that this is different to Ref. \cite{YNT+03} which claimed $L_{\varpi}$ was conserved.

\subsection{Semi-classical fixed points}

The equations of motion for the classical analogue of the $E\otimes\varepsilon$ system are similar to those of the $E\otimes\beta$ (\ref{ceq1}-\ref{ceql}), except now there is an extra degree of freedom from the additional oscillator mode. For Hamiltonian \eqrf{englman::hamiltonian}, there exists two fixed points at the origin position of the oscillators, with $L_\varpi=\pm 1$, and then a ring of stable fixed points around the origin, satisfying $L_{\varepsilon}^{2}+L_{\theta}^{2} = 1$, with $q_{\varepsilon} = -\frac{L}{2\sqrt{\omega}},L_{\varepsilon},q_{\theta}  = \frac{L}{2\sqrt{\omega}}$. Note the correspondence to the potential, figure \ref{mh}.

\subsection{Ground State Ansatz}

From the work of Englman \cite{Eng62,Eng72}, we now introduce the following ansatz for the ground state of the Hamiltonian \eqrf{englman::hamiltonian}. This approximation is based on a similar construction to that of the eigenstates in the $E\otimes\beta$ case.
\begin{equation}
\label{englman::ansatz}
\braket{q,\phi}{\Psi} = \frac{1}{\sqrt{2}} e^{-L^2/(2\hbar\omega)^2} \left( A(q,\phi) |\da\rangle - iB(q,\phi) |\ua\rangle \right)
\end{equation}
where
\bea
    A(q,\phi) & = & e^{-q^2/2}\left(\cosh\left(\frac{qL}{2\hbar\omega}\right) + e^{i\phi} \sinh\left(\frac{qL}{2\hbar\omega}\right)\right) \nonumber\\
    B(q,\phi) & = & e^{-q^2/2}\left(\cosh\left(\frac{qL}{2\hbar\omega}\right) - e^{i\phi} \sinh\left(\frac{qL}{2\hbar\omega}\right)\right)\nonumber
\eea
and we have adopted a polar coordinate system for the oscillator variables $q_\theta = q\cos(\phi)$, $q_\epsilon = q\sin(\phi)$. Note that $\phi$ commutes with $q$. The ground state is degenerate, and the orthogonal ground state to $\ket{\Psi}$ is simply its complex conjugate, $\ket{\Psi^*}$, i.e., $\braket{\Psi}{\Psi^*}=0$.

It was shown in \cite{Eng62,Eng72} that this ansatz gave a good approximation to the ground state energies of the Hamiltonian \eqrf{englman::hamiltonian}. In this section we shall use it to derive an expression for the ground state spin-oscillators entanglement. We find good agreement between this expression and numerical results.

Entanglement between the spin and the two oscillators can be calculated from the von Neumann entropy of the spin's reduced density matrix obtained by taking the partial trace over the oscillator variables.
\begin{equation}
    \rho_s = \int_0^{2\pi} \int_0^\infty \ket{\Psi(q,\phi)} \bra{\Psi(q,\phi)}\,q\,d\phi\,dq
\end{equation}

In calculating $\rho_s$ we will make much use of the following integrals
\begin{equation}
\int_0^{\infty} e^{-q^2} \cosh^2(\alpha q)\,q \,dq
    = \frac{1}{4}\left(2+  e^{\alpha^2} \alpha\sqrt{\pi}\mathrm{Erf}(\alpha) \right) \nonumber
\end{equation}
\begin{equation}
    \int_0^{\infty} e^{-q^2} \sinh^2(\alpha q)\,q \,dq  = \frac{1}{4} e^{\alpha^2} \alpha\sqrt{\pi}\mathrm{Erf}(\alpha)\nonumber
\end{equation}
\begin{equation}
    \int_0^{\infty} e^{-q^2} \cosh(\alpha q) \sinh(\alpha q)\,q \,dq  =  \frac{1}{4} e^{\alpha^2} \alpha \sqrt{\pi}\nonumber
\end{equation}
where $\mathrm{Erf(x)}$ is the error function ranging between $0$ and $1$. We will further require the integrals
\begin{equation} \label{eqn::ABst}
\int_0^{2\pi} \int_0^\infty A(q,\phi) B(q,\phi)^*\, d\phi\,q\,dq = \pi\nonumber
\end{equation}
The state $\braket{q,\theta}{\Psi)}$ is not normalized. Its normalization factor $N$ is given by
\bea
    N^2 & = & \langle\Psi(q,\phi)|\Psi(q,\phi)\rangle \nonumber \\
                & = & \frac{1}{2}e^{-2L^2/(2\hbar\omega)^2} \nonumber \\
                &   & \times \int_0^{2\pi} \int_0^\infty |A(q,\phi)|^2+|B(q,\phi)|^2\,d\phi\,q\,dq \nonumber \\
                & = & \pi e^{-2L^2/(2\hbar\omega)^2} \left[ 1 + e^{L^2/(2\hbar\omega)^2} \frac{L\sqrt{\pi}}{2\hbar\omega}
                        \mathrm{Erf}\left(\frac{L}{2\hbar\omega}\right) \right] \nonumber
\eea

The reduced density matrix of the spin system is then
\bea
    \rho^S & = & \frac{e^{-2L^2/(2\hbar\omega)^2}}{2N^2} \int_0^{2\pi} \int_0^\infty \Big( |A(q,\phi)|^2  \ket{\da}\bra{\da} \nonumber \\
            &   & + A(q,\phi)B(q,\phi)^* \ket{\da}\bra{\ua} \nonumber \\
            &   & + A(q,\phi)^* B(q,\phi) \ket{\ua}\bra{\da} \nonumber \\
            &   & + |B(q,\phi)|^2 \ket{\ua}\bra{\ua} \Big) \,d\phi\,q\,dq
\eea
and using the above integrals this evaluates to
\begin{equation}
\rho^S = \frac{1}{2} \left[ \begin{array}{cc}
    1 & i \mathcal{C}(L/\omega)   \\
    -i \mathcal{C}(L/\omega)  & 1
\end{array} \right] \\
\end{equation}
where $\mathcal{C}(L/\omega) =  \left[1+ e^{L^2/(2\hbar\omega)^2} \frac{L\sqrt{\pi}}{2\hbar\omega} \mathrm{Erf}\left(\frac{L}{2\hbar\omega}\right) \right]^{-1}$. It can be readily seen for large coupling $L/\omega \rightarrow \infty$ we have $\mathcal{C}(L/\omega) \rightarrow 0$ and the state $\rho^S$ is completely mixed. The entanglement of formation between the spin and the oscillators, given by $S(\rho^S)$ takes its maximum value of $1$ in the strong coupling limit. On the other hand for small coupling $L/\omega << 1$, we find $\mathcal{C}(L/\omega)$ is also close to one and $\rho_S$ approaches a pure state and the entanglement of formation for the system approaches zero.

The reduced density matrix for the orthogonal degenerate ground state $|\Psi^*\rangle$ is simply the adjoint of $\rho^S$ and its entanglement properties are identical. Somewhat surprisingly, however, a ground state superposition of these two displays different entanglement properties.

Consider an arbitrary such superposition
\begin{equation}
    c_1 \braket{q,\phi}{\Psi} + c_2  e^{i\gamma} \braket{q,\phi}{\Psi^*}
\end{equation}
where $c_1^2 + c_2^2 = 1$. Neglecting normalization for the moment the density matrix entries for the system can be written in the $\ket{\da}, \ket{\ua}$ basis as
\bea
\rho_{00}(q,\phi) & = &  \left| c_1 A(q,\phi) + c_2 e^{i\gamma} A(q,\phi)^* \right|^2\nonumber \\
\rho_{01}(q,\phi) & = & i \left(c_1 A(q,\phi) + c_2 e^{i\gamma} A(q,\phi)^* \right) \nonumber \\
    & & \times \left(c_1 B(q,\phi)^* - c_2 e^{-i\gamma} B(q,\phi) \right) \nonumber\\
\rho_{10}(q,\phi) & = & -i \left(c_1 A(q,\phi)^* + c_2 e^{-i\phi} A(q,\gamma) \right) \nonumber \\
    & & \times \left(c_1 B(q,\phi) - c_2 e^{i\phi} B(q,\gamma)^* \right) \nonumber\\
\rho_{11}(q,\phi) & = & \left| c_1 B(q,\phi) - c_2 e^{i\gamma} B(q,\phi)^* \right|^2 \nonumber
\eea
and, as before, we can calculate the reduced density matrix entries
\bea
\label{superpos00}
\rho^S_{00} & = & \int_0^{2\pi} \int_{-\infty}^\infty \rho_{00}(q,\phi)\,d\phi\,dq  \nonumber \\
    & = & (c_1^2+c_2^2) \int_0^{2\pi} \int_{-\infty}^\infty |A(q,\phi)|^2 \,d\phi\,q\,dq \nonumber \\
    & + & 2c_1c_2 \int_0^{2\pi}  \int_{-\infty}^\infty \mathrm{Re}\left( e^{-i\gamma} A(q,\phi)^2 \right) \,d\phi\,q\,dq \nonumber \\
\eea
The first term we have already calculated and we find
\begin{multline}
  \int_0^{2\pi} \int_0^\infty \mathrm{Re}\left(e^{-i\gamma} A(q,\phi)^2 \right) \,d\phi\,q\,dq \nonumber \\
       = \frac{\pi}{2} \cos(\gamma)\left( 1 + \left[1 + e^{L^2/(2\hbar\omega)^2} \frac{L\sqrt{\pi}}{2\hbar\omega} \mathrm{Erf}\left(\frac{L}{2\hbar\omega}\right) \right] \right)\nonumber
\end{multline}
Reintroducing the normalization factor into equation \ref{superpos00} gives us
\begin{equation}
    \rho^S_{00} = \frac{1}{2} \left( 1 + c_1 c_2 \cos(\gamma)\left(1+\mathcal{C}(L/w)\right) \right)
\end{equation}
with $\mathcal{C}(L/w)$ as before. A similar calculation finds
\begin{multline}
   \rho^S_{01} = \frac{i}{2} \left(c_1^2 - c_2^2\right) \mathcal{C}(L/w) \\
 - \frac{1}{2} c_1 c_2 \sin(\gamma) \left( 1 + \mathcal{C}(L/w) \right)
\end{multline}
and since $\rho^S$ is a density matrix the remaining two entries are $\rho^S_{10} = \left(\rho^S_{01})\right)^*$, $\rho^S_{11} = 1- \rho^S_{00}$.

The eigenvalues of $\rho_S$ can be written as $\frac{1}{2}\left(1 \pm \sqrt{1-\Gamma}\right)$ where
\begin{equation}
    \Gamma = 1 - c_1^2 c_2^2\left(1 + \mathcal{C}(L/\omega)\right)^2 - \left(c_1^2 - c_2^2)\right)^2 \mathcal{C}(L/\omega)^2
\end{equation}
Interestingly we see that if we take an equal superposition $c_1 = c_2 = \frac{1}{\sqrt{2}}$ then let the coupling become very strong $L/\omega \rightarrow \infty$ these eigenvalues become $\frac{1}{4}, \frac{3}{4}$ and the entanglement of formation is $S(\rho^S) \approx 0.8113$. The spin-oscillator entanglement in this ground state can never reach a maximum value regardless of how large the coupling term is. This is quite different from the corresponding results in the $E\otimes \beta$ model, where there exists a ground state separable for all couplings. This will be discussed further in Sec. \ref{ES}.

\begin{figure}[ht]
\begin{center}
\scalebox{.45}{\includegraphics{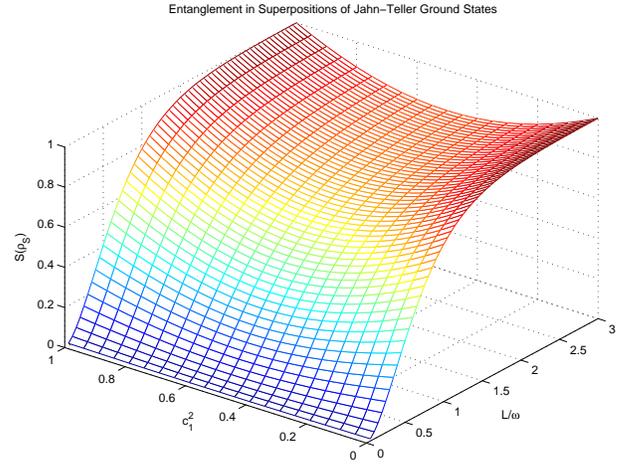}}
\end{center}
\caption{The entanglement in the (ansatz) ground state of the $E\otimes\varepsilon$ model with respect to the coupling strength for all possible superpositions of the two degenerate states. Note that this is quite different from the corresponding results for the $E\otimes\beta$ model (shown in figure \ref{qbosc}). Here the ground state is \emph{always} entangled, regardless of the superposition.}
\label{cir}
\end{figure}

\subsection{Numerical analysis}

We now compare the ground state entanglement results from the ansatz with exact numerical results. The Hilbert spaces of the two oscillators were truncated to $50$ basis states. Increasing the Hilbert space further had no effect on the results.

\begin{figure}[ht]
\begin{center}
\scalebox{.45}{\includegraphics{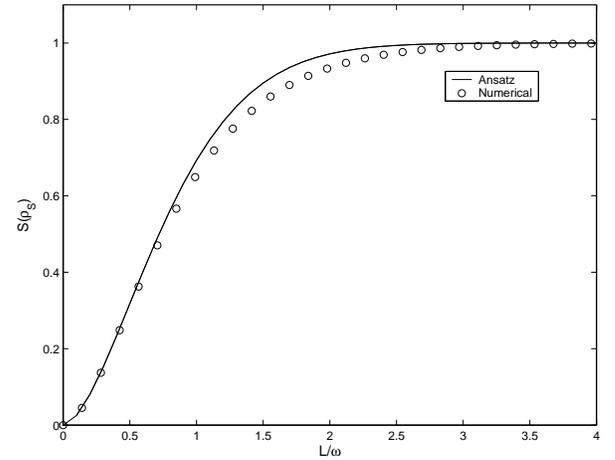}}
\end{center}
\caption{Comparison of the entanglement in the ground state between the exact results obtained by numerical diagonalization and the ansatz of Eq. \eqrf{englman::ansatz}.}
\label{compeval}
\end{figure}

\begin{figure}[ht]
\begin{center}
\scalebox{.45}{\includegraphics{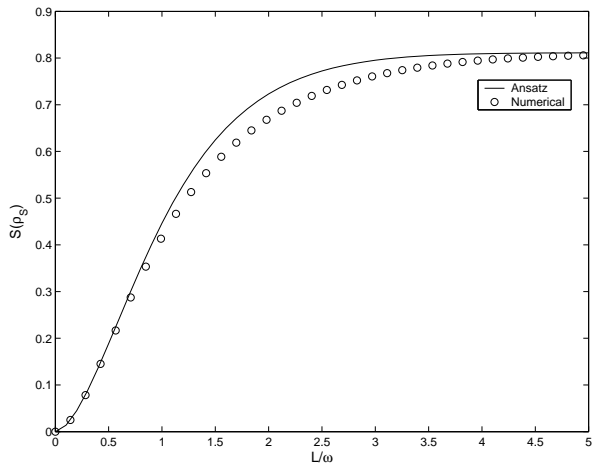}}
\end{center}
\caption{Comparison of the entanglement in the ground state between the numerical diagonalization and the ansatz for an equal superposition of the two orthogonal states.}
\label{compent}
\end{figure}

We see that there is good agreement between the exact numerics and the ansatz, particularly in the small and large coupling limits.

\subsection{Distributed Entanglement}\label{ES}

The entanglement in the ground state we have considered so far is that between the qubit and the pair of oscillators. Since the oscillators, and the couplings to the qubit, are identical, the entanglement to the qubit is distributed equally between the two oscillators. However, it is possible to consider quantitatively how the entanglement between the qubit and the oscillators is shared between the two polar degrees of freedom - radial and angular coordinates. Such entanglement involving (orthogonal) internal degrees of freedom, as opposed to physical partitions of the system have been considered, for example, in the context of trapped ions \cite{MMK+03}, where spin and orbital degrees of freedom of a single ion are entangled.

In the limit of large coupling, $L/\omega \gg 1$, the ground state ansatz can be expressed as
\begin{equation}\label{sepr}
\braket{q,\phi}{\Psi} \approx \mathcal{F}(q) \left(\ket{0} - i\ket{1} + e^{i\phi}\left(\ket{0}+i\ket{1}\right)\right)
\end{equation}
where $\mathcal{F}(q)=e^{(-L^2/(2\hbar\omega)^2} e^{(qL/2\hbar\omega) - q^2/2}/(\sqrt{2}N)$. The radial coordinate, $q$, is separable, hence the qubit is entangled solely with the angular degree of freedom, $\phi$ of the two oscillators. Outside of this parameter range however, the radial coordinate is not separable, meaning the qubit is entangled with both degrees of freedom.

To quantify this distribution of entanglement for the ground state \eqrf{englman::ansatz}, it is possible to determine the entanglement solely between the angular degree of freedom and the qubit. We begin by identifying the states
\begin{equation}
\mathcal{U}_{m}(\phi) = \frac{1}{\sqrt{2\pi}}e^{\pm im\phi}
\end{equation}
as eigenstates of $\op{L}_\varpi = i\hbar\partial_{\phi}$, with eigenvalue $m$. In the ground state, $\ket{\psi}$, only the $m=0,1$ states are present. So the angular degree of freedom, $\phi$, is constrained to a two-dimensional subspace of its total Hilbert space. Letting $\ket{0}_\phi \equiv \mathcal{U}_{0}(\phi)$ and $\ket{1}_\phi\equiv\mathcal{U}_{1}(\phi)$, we may view the angular degree of freedom in the ground state as itself a qubit, reducing the problem of the entanglement between the ({\it spin}) qubit and $\phi$, to the well-known situation of a pair of qubits.
Rewriting the state of the({\it spin}) qubit in the basis $\ket{+} = \left(\ket{\downarrow}+i\ket{\uparrow}\right)/\sqrt{2}$ $\ket{-} = \left(\ket{\downarrow}-i\ket{\uparrow}\right)/\sqrt{2}$ (which are the eigenstates of $\sigma_{\varpi}$), the ground state in the limit of large coupling, Eq. \eqrf{sepr} becomes
\begin{equation}
\braket{q}{\Psi} \approx \sqrt{2}\mathcal{F}(q) \left(\ket{-}\ket{0}_\phi +\ket{+}\ket{1}_\phi\right)
\end{equation}

Clearly, the spin-qubit and the '$\phi$'-qubit are in a maximally entangled Bell state, completely separable from the radial coordinate.

 \begin{figure}[t!]
\begin{center}
\scalebox{.45}{\includegraphics{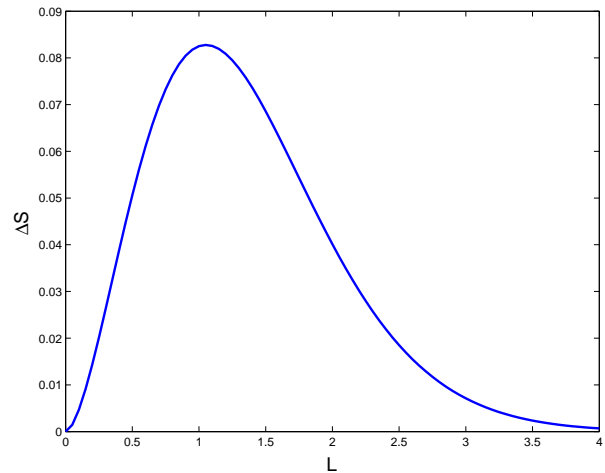}}
\end{center}
\caption{The difference between the qubit-oscillators entanglement and the qubit-angular degree of freedom entanglement (in terms of the von Neumann entropy) as a function of $L/\omega$. Hence, almost all the entanglement is between the qubit and the angular degree of freedom.}
\label{conc}
\end{figure}

The {\it concurrence} \cite{HW97,Woo01} is a good measure of the two-qubit mixed-state entanglement, which we can use to quantify the entanglement between the spin-qubit and the $\phi$-qubit.

The concurrence, $C$, between a pair of qubits, $A$ and $B$ is defined using the ``spin-flipped'' density matrix
\begin{equation}
\tilde{\rho}_{AB} = \left(\sy \otimes \sy\right)\rho^*_{AB} \left(\sy \otimes \sy\right)
\end{equation}
where the asterisk is the complex conjugation in the standard basis. If the square roots of the eigenvalues of the product $\rho_{AB}\tilde{\rho}_{AB}$ in decreasing order are $L_1,L_2,L_3,L_4$, then the concurrence of the density matrix $\rho_{AB}$ is
\begin{equation}
C = \textrm{min} \left\{0,L_1-L_2-L_3-L_4\right\}.
\end{equation}

The concurrence is related to the von Neumann entropy via the tangle, $\tau = C^2$, by
$$
S = \mathbb{H}\left(\frac{1 + \sqrt{1-\tau}}{2}\right)
$$
where $\mathbb{H}(x) = -x\log_2(x) - (1-x)\log_2(1-x)$ is the Shannon entropy.

Using the above, it is possible to calculate the entropy of entanglement between the qubit and the angular degree of freedom, and compare it to the total entanglement between the qubit and the two oscillators. Figure \ref{conc} shows the difference, $\Delta S$, between these two entanglements. $\Delta S$ asymptotes to zero, such that in the strong coupling regime, the qubit becomes disentangled from the radial degree of freedom and is solely entangled with the angular degree of freedom, as predicted by Eq.\eqrf{sepr}. Furthermore, this entanglement is maximal. Note that $\Delta S$ is relatively small implying that qubit-oscillators entanglement is concentrated between the angular degree of freedom and the qubit.

The two orthogonal degenerate ground states, $\ket{\Psi}$, and it's complex conjugate, $\ket{\Psi^*}$ from the ansatz Eq. \eqrf{englman::ansatz}, are the two sole ground states where the angular degree of freedom can be treated analogous to a qubit. In any superposition of these two states, the states of the angular degree of freedom is in the subspace spanned by the states $\mathcal{U}_m(\phi)$ with $m = 0,\pm 1$ - now a three-level system, or {\it qutrit}.

As shown in figure \ref{cir}, for any superposition, the ground state entanglement does not asymptote to the maximal value, but however, there is no ground state superposition that has zero entanglement, as in the single oscillator ($E\otimes \beta$) case. In the large coupling limit, the radial degree of freedom still becomes separable, such that the entanglement is concentrated between the qubit and the angular degree of freedom for all ground possible states. The observation that in all superpositions the angular degree of freedom is viewed as a qutrit rather than a qubit could explain why the entanglement is never zero - as seen in the single oscillator case - nor maximal.

\subsection{Addition of transverse magnetic field}

For completeness, we now consider the effect of applying a transverse magnetic field to the qubit, in the direction perpendicular to the oscillator displacements. The Hamiltonian thus becomes
\begin{equation}
H = \Delta\op{\sigma}_\varpi + \hf\omega\left(p_{\varepsilon}^{2} + q_{\varepsilon}^{2} + p_{\theta}^{2} + q_{\theta}^2 \right) + \hf L\left(q_{\theta}\st + q_{\varepsilon}\hat{\sigma}_{\varepsilon}\right)
\end{equation}
where $\Delta$ is the strength of the magnetic field.

With respect to the fixed point structure in the classical analogue, the addition of the $\Delta$ term has the effect of destroying the stable ring of fixed points, leaving four stable points, at
\[
L_{\varepsilon} = \pm\sqrt{1 - \frac{16\omega^2\Delta^2}{L^4}},L_{\varpi} = -\frac{4\omega\Delta}{L^2}\]
and
\[
L_{\theta} =\pm\sqrt{1 - \frac{16\omega^2\Delta^2}{L^4}},L_{\varpi} = -\frac{4\omega\Delta}{L^2}.
\]
This implies that again a pitchfork bifurcation is present, with the critical coupling, $L^2 = 16\omega^2\Delta^2$. This should again manifest itself in the large $\Delta$ limit of the entanglement in the ground state, as in the $E\otimes\beta$ model.

Moving to the quantum regime, the ground state is no longer degenerate, and the presence of the transverse field forces the ground state into a maximally entangled state in the large coupling limit.

\begin{figure}[ht]
\begin{center}
\scalebox{.5}{\includegraphics{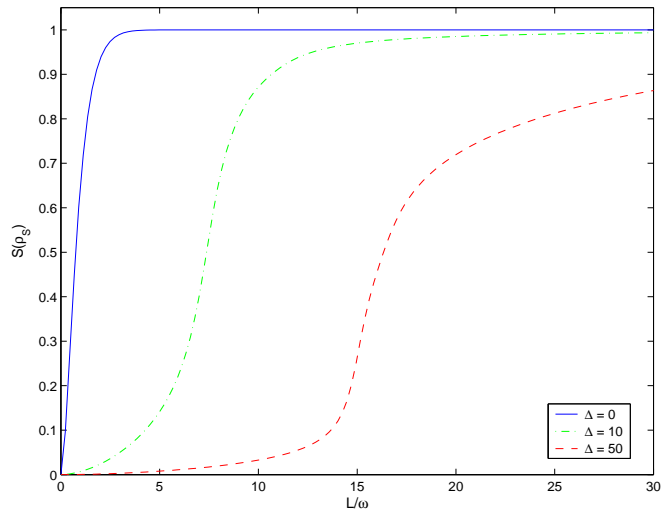}}
\end{center}
\caption{Entanglement in the ground state of the $E \otimes \varepsilon$ model in the presence of a transverse magnetic field. Note the similarities with the $E\otimes\beta$ model results in figure \ref{entsosc}.}
\label{withd}
\end{figure}

From figure \ref{withd}, it's clear that the bifurcation in the $E\otimes\varepsilon$ model plays a similar role as that in the $E\otimes\beta$ model, in the large $\Delta/\omega$ limit (the classical limit of the oscillator). The gradient of the entropy of entanglement curve with respect to $L$ becomes steeper around the critical point, and we see the division into the ``separable'' and ``entangled'' parameter regions.

\section{Conclusion}

We have studied the entanglement in the ground states of the $E\otimes \beta$ and the $E\otimes\varepsilon$ Jahn-Teller systems, which model a single qubit coupled to one and two harmonic oscillators, respectively.

In the single oscillator case, we have considered the results of both Levine and Muthukumar \cite{LM03} and Lambert {\it et al.} \cite{LEB03}. In both cases, we have argued that the entanglement characteristics of the ground state can be understood by considering the bifurcation of the fixed points in the classical counterpart. In the two extremes considered in \cite{LM03} and \cite{LEB03} the classical limit becomes relevant - either of the oscillator, or the entire system, respectively. Again, as shown in previous work \cite{HMM03a}, the nature of the bifurcation (the {\it pitchfork} structure) is crucial - a single fixed point becomes two, leading to superposition states in the quantum regime.

In the $E\otimes\varepsilon$ model, we found that the ground state entanglement between the qubit and the oscillators differed from that of the single oscillator model, insofar as for no superposition of the orthogonal ground states was there zero entanglement. Furthermore, how this entanglement is shared between the two degrees of freedom of the double oscillator subsystem was considered. It was found that the entanglement between the qubit and the two oscillators is concentrated between the qubit and the angular coordinate, with the radial coordinate becoming completely separable in the large coupling limit. This correlation between the angular degree of freedom and the qubit states is not surprising given the radial symmetry of the potential created by the qubit-oscillators coupling.

The Hamiltonian of the $E \otimes\varepsilon$ model  in Eq. \eqrf{var_englman} can be
physically realized using two vibrational degrees of freedom of a
single trapped ion \cite{STK97}. The required coupling is achieved
using external laser pulses to couple different components of the
atomic polarization vector, $\vec{\sigma}$ to each of the vibrational
modes.

In Ref. \cite{Sjo00}, Sj\"ovist used the $E\otimes\varepsilon$ Jahn-Teller system as a model for electron nuclear interaction. While the entanglement in higher energy eigenstates was considered in that article, our results for the ground state in the large coupling limit coincide. Our results will hopefully shed more light on the characteristics of this electron-nuclear entanglement in molecular ground states.

One of the most intriguing results of this paper
is that the $ E \otimes \varepsilon$ Jahn-Teller
model always has an entangled
ground state and when we take the semi-classical
limit ($L/\omega \to \infty$) the entanglement
between the qubit and the oscillators approaches
its maximal value. In contrast, for the $ E \otimes \beta$
model there are two degenerate ground states for which
there is no entanglement. It appears that this
difference is due to the presence of the angular
degree of freedom for the oscillators.
We conjecture that the entanglement is intimately
connected with the geometric (Berry's) phase associated
with cyclic adiabatic variations of
the angular co-ordinate of the classical limit of this model \cite{Mea92}.

The above raises an important question as
to whether our results are a manifestation
of a very general phenomena connecting entanglement
and geometric phases. In the hope of
stimulating further work we offer the following conjecture.

{\bf Conjecture:}
{\it Let $H(S)$ be a Hamiltonian
which depends on some
parameter $S$ and acts on a bipartite Hilbert
space $V = V_Q \otimes V_C$ of finite dimension.
 Suppose that in some limit $S \ra S_{cl}$,
the Hamiltonian becomes $H(C)$ which acts on the Hilbert space $V_Q$
where $C$ denotes a finite dimensional parameter.
Suppose also that there is geometric phase associated with
cyclic adiabatic variations of $C$.
Then for all possible ground states of
$H(S)$ there is always
entanglement between $V_Q$ and $V_C$.
Furthermore, the entanglement approaches
its maximum possible value  as $S \ra S_{cl}$.}

This conjecture should first be tested for the $T \otimes H$ Jahn-Teller model which describes three-fold degenerate electronic levels coupled to a five-fold degenerate phonon and which is relevant to fullerene (C$_{60}$) molecules \cite{HG00}.\\


\acknowledgements
This research is supported by the Australian Research Council as part of the Centre of Excellence for Quantum Computer Technology.

\appendix
\section{Numerical basis - displaced Fock states}\label{basis}

Numerical analysis of a system within an infinitely dimensional space often implies some truncation of the Hilbert space for calculations. For the $E\otimes\beta$ system, to reduce the potential numerical error from this truncation, rather than choosing the set of Fock states as the basis for the Hilbert space of the oscillator, we use the displaced Fock states \eqrf{gsb1},\eqrf{gsb2} corresponding to the eigenstates for $\Delta = 0$. In this basis, the Hamiltonian is diagonal for $\Delta = 0$, with entries given by the energy eigenvalues \eqrf{energies}. For, $\Delta \neq 0 $, we must calculate the off-diagonal elements.
Since this set of basis states is not orthogonal, to determine the off-diagonal elements of the Hamiltonian matrix we make use of the following expressions:
\begin{itemize}
\item $\braket{\psi_{m}^L}{\psi_{n}^L} = \braket{\psi_{m}^R}{\psi_{n}^R}=\delta_{mn}$,
\item $\braket{\psi_{m}^R}{\psi_{n}^L} = \braket{\psi_{m}^L}{\psi_{n}^R}=0$,
\item $\int \chi_m^L(q) \chi_n^L(q) dq = \delta_{mn}$,
\item $\int \chi_m^R(q) \chi_n^R(q) dq = \delta_{mn}$,
\item $\int \chi_m^L(q) \chi_n^R(q) dq = \bra{m}\op{D}(2L / \omega^2)\ket{n}$,
\item $\int \chi_m^R(q) \chi_n^L(q) dq = \bra{m}\op{D}(-2L / \omega^2)\ket{n}$,
\end{itemize}
where $\op{D}(\alpha)$ is the displacement operator. From Caves {\it et al.} \cite{CTD+80} we have
\begin{equation*}
\bra{m}\op{D}(\beta)\ket{n} = \left\{ \begin{array}{ll}
  \displaystyle\frac{\sqrt{\frac{n!}{m!}}\beta^{m-n} L_n^{(m-n)}\left( |\beta|^2\right)}{e^{\frac{|\beta|^2}{2}}} & \textrm{if $m>n$}\\
&\\
\displaystyle\frac{\sqrt{\frac{m!}{n!}}(\beta^*)^{n-m} L_n^{(n-m)}\left(|\beta|^2\right)
}{(-1)^{m+n}e^{\frac{|\beta|^2}{2}}} & \textrm{if $n\geq m$}
\end{array}\right.
\end{equation*}
where $L_r^s (u)$ are the generalized Laguerre polynomials. Now, considering $N+1$ oscillator modes, the non-zero off-diagonal elements of Hamiltonian \eqrf{hamqbosc} matrix, $\mathcal{H}$  are given by
\begin{equation*}
\begin{split}
\mathcal{H}(m+1,n+N+2)  &= \bra{\psi^L_m}H\ket{\psi^{R}_n} \\
                        &= \Delta \braket{\chi_m^L}{\chi^R_n}\bra{\uparrow}\sx\ket{\downarrow}\\
                        &= \Delta \bra{m}\op{D}^{\dagger}\left(-L/\omega^2\right) \op{D}\left(L/\omega^2\right)\ket{n} \\
                        &= \Delta\bra{m}\op{D}\left(2L/\omega^2 \right)\ket{n}
\end{split}
\end{equation*}
and similarly,
\begin{equation*}
\mathcal{H}(m+N+2,n+1) = \Delta\bra{m}\op{D}\left(-2L/\omega^2 \right)\ket{n},
\end{equation*}
both of which can be evaluated using the expressions above.

\bibliography{JahnTeller}

\end{document}